# Did the Corona Borealis/A2142 supercluster binary-like system originate as a proto-cluster binary embedded in a primordial cloud of galaxies?

**Giovanni C. Baiesi Pillastrini**[1*]

## ABSTRACT

The formation of the giant binary-like system composed by the Corona Borealis and Abell 2142 superclusters is an intriguing conundrum of the formation of large scale structures since, from the observational point of view, it represents a rare peculiarity in the distribution of massive galaxy superclusters. Having a configuration similar to a giant binary system interconnected by a huge filament, it is likely one if not the unique case to date observed in the Local Universe. So a question arise: how and when did it form? Here, the CrB/A2142 system has been hypothesized to be a descendent of a primordial binary system composed of close galaxy proto-clusters originated from two independent collapsing processes within a dense cloud of galaxies. Then, due to the interplay of gravity-antigravity (DE), at a certain stage of the binary evolution, they decoupled and began to move away in radial motion from each other following the accelerate expansion of the Universe. To define the look-back time at which the event of decoupling occurred, a Newtonian approximation of the two-body motion in presence of DE has been applied using current physical parameters of the CrB/A2142 system. Its compatibility with the observed era of formation of primordial galaxy proto-clusters has been tested. The event of the binary decoupling has been established at $14.7^{+3.6}_{-2.2}$ *Gys* ago, a time larger than the current age of the Universe. Therefore, unless the Universe is older than believed, the suggested hypothesis on the origins of the CrB/A2142 system should be rejected even if the error lower limit of the calculated look-back time is ~ 12.5 *Gys* ( $z$ ~ 4.5) largely consistent with observations.

[1] *Independent researcher*
[*] *permanent address: via Pizzardi, 13 - 40138 Bologna - Italy - email: gcbp@micso.net*

## 1. Introduction

This study is the sequel of two previous papers devoted to demonstrating the binary origin of two well-known galaxy superclusters of our local Universe: the Corona Borealis (CrB) and A2142 (Baiesi Pillastrini, 2016, 2019, BP16/19). Their close proximity interconnected by a huge filament of galaxy groups and clusters approximates a detached binary-like system (CrB/A2142sys). The rarity of its peculiar configuration represents a novelty in the observed distribution of large scale structures. In fact, the observed superclusters are mostly *isolated* systems defined either "basin of attraction" (Valade et al. 2024) or "supercluster cocoon" (Einasto et al. 2021) suggesting the idea that they form and grow through merging processes. The distribution of peculiar velocities belonging to many clusters of the CrB/A2142sys shows signatures of mutual gravitational interactions characterized and a past matter flow toward the Corona Borealis supercluster even if this trend cannot be taken seriously because of the too large uncertainties affecting the peculiar velocity measurements (BP19). Besides, the CrB/A2142sys shows a substantial dynamical equilibrium. In fact, the analysis of the reciprocal tidal influences confirms that inner dynamics of both superclusters remain unperturbed by the actions of external tidal forces at least within the spherical surfaces delimited by respective turnaround radii. Outskirts alone are subject to moderate tidal influences (BP19). Gravitational interactions between the Corona Borealis and A2142 superclusters were firstly advanced by Bahcall and Soneira (1984) adding the prediction that they are likely evolving to a future collapse to form a single, extended structure (see also Luparello et al. 2011). Predictions recently confirmed by Sankhyayan et al. (2023) that identifies the CrB/A2142sys as a single supercluster dubbed as the SCL2. Instead, as stated before by BP19, unlikely the CrB/A2142sys is a bound structure and/or could collapse in single superstructure unless the CrB is much more massive than A2142. Anyway, if the future evolution of the CrB/A2142sys could be predicted from available data, nothing is known about its origin and past evolution. How and when did it form? Here, we advance a possible answer hypothesizing that the CrB/A2142sys is a *descendent* of a primordial binary system composed of two proto-clusters originated from two independent collapsing process within a dense cloud of galaxies. An idea that would be supported by two recent and independent studies: first by Nicandro Rosenthal et al. (2025) that confirmed the presence of a massive galaxy proto-cluster at z~3.14. The peculiarity of this object is the presence of two overdensities at $z = 3.1$ and $z = 3.14$ (GNCL-z3.10 and GNCL-z3.14) respectively, that are embedded in its volume , a sort of a "Two-in-One" system. Using an analytical spherical collapse model, they calculated that both overdensities should collapse and virialize at $0.5 < z < 0.8$ with masses similar to that of the Virgo cluster at $z$=0. Besides, the study predicts a long merging process of the two substructures which will end in the formation of an unique structure expected to collapse at $0.1 < z < 0.4$ in a single massive supercluster of ~ $10^{15}$ $M_{Sun}$ . Second, Sankhyayan et al. 2023 (S23) published a new Supercluster Catalog based on the SDSS survey and the WHL 12/15 Cluster catalog of Wen et al. (2012, 2015) where 662 superclusters were identified using a modified Friend of Friend algorithm in the redshift range of $0.05 \leq z \leq 0.42$. Within the Catalog, the CrB/A2142sys has been identified as the SCL2, a single supercluster at $z$~ 0.079 with a size of 124 *Mpc* and a total mass of ~ 2.55 x $10^{16}$ $M_{Sun}$. In their Fig. 8, a fair 3D representation of the SCL2 clearly shows the CrB and A2142 superclusters and its interconnecting filament. Curiously, both superclusters appear as two dominant overdensities within the SCL2 (again a Two-in-One configuration). Intriguingly, both these studies seem to support our advanced hypothesis for the formation of the CrB/A2142sys within a dense cloud of galaxies. A *test of compatibility* has been set up to verify the necessary condition (even if not sufficient) that the hypothetical decoupling of the CrB/A2142sys happened during the observed era of formation of the primordial galaxy proto-clusters.
The assumed cosmological parameters are *$H_0$ = 68 $km\ s^{-1}\ Mpc^{-1}$, $\Omega_m = 0.3$ and $\Omega_\Lambda = 0.7$.
The paper is organized as follows: in Sect. 2, the past history of the CrB/A2142sys has been analyzed using an approximated Newtonian two-body problem; in Sect.3, results are drawn and in Sect. 4, the summary. .

## 2. Setting up the test of compatibility in presence of DE (**Λ=const.)

The complex object found by Nicandro Rosenthal et al. (the Two-in-One) has been assumed as a possible progenitor model of formation for the CrB/A2142sys. However, contrary to their conclusion for which that object will end its evolution with the formation of a single massive structure, from BP19 we know that the CrB/A2142sys unlikely will collapse in single bound structure. So that, its present configuration could be assumed as a *detached* binary system where signatures of gravitational interactions and their connecting filament of galaxy groups and clusters are remnants of the past evolution of a primordial binary system driven by the interplay between gravity and anti-gravity (of DE).

* $H_0 = H_\Lambda = \sqrt{\frac{8\pi}{3} G\rho_\Lambda}$ which is the effective asymptotic Hubble constant in ΛCDM cosmology that is, $H_0 \rightarrow H_\Lambda$ at $t \rightarrow \infty$.

** Recent surveys of the DESI Collaboration found that *DE changes over time* (Abdul-Karim et al. 2025, Lodha et al. 2025) which, if confirmed, should influence either the Universe's expansion at different points in its history and the formation and evolution of structures like galaxy superclusters.(Pfeifer et al. 2020). In that case an exact solution for the present two-body problem cannot be yet obtained.

### 2.1. When the CrB/A2142sys decoupled?

The fundamental parameter to be defined for the test is *when* the CrB/A2142sys broke its gravitational binding and then began to expand. With the finding of Dark Energy as cause of the accelerated expansion of the Universe (Schmidt et al. 1998, Perlmutter et al. 1999) and its impact on the dynamics of astronomical objects, many studies concerning the two-body problem in presence of the dark energy have been published, among others, Emelyanov & Kovalyov (2013), Emelyanov et al.(2015), Bisnovatyi-Kogan & Merafina (2019, 2023), McLeod & Lahav, 2020) and Benisty et al. (2024). An exelent treatment of this topic can be found in Silbergleit & Chernin ( 2019).

#### 2.1.1. Some assumptions

1) The SCL2 (the supercluster found by S23 in which the CrB/A2142sys is embedded) has gravitational influence on the dynamics of the CrB/A2142sys only if it is gravitationaly *bound*. Now, in the framework of the ΛCDM cosmological model, the gravitational potential of a clustered structure is given by $\Phi = -\Phi_g - \Phi_\Lambda$ where $\Phi_g$ is the *attractive component* of the potential due to gravity and $\Phi_\Lambda$ is the *repulsive component* due to dark energy. They are given respectively by

$$\Phi_g = \frac{GM}{R} \qquad \text{and} \qquad \Phi_\Lambda = \frac{4}{3} G\rho_\Lambda R^2$$

where $G$ is the gravitational constant and $\rho_\Lambda$ is the dark energy density of $6x10^{-30} g\ cm^{-3}$ (Plank collaboration 2015, Prat et al. 2022). Then, if the SCL2 is gravitationally bound , the inequality $|2\Phi_\Lambda| < |\Phi_g|$ must be satisfied (Chernin et al. 2012). To verify it the size and total mass of the SCL2 has been taken from S23. After correction for the different cosmological parametrization, the size is 133.6 $Mpc \rightarrow R \approx 67\ Mpc$ and $M = 2.7 \text{ x } 10^{16}\ M_{Sun}$ . The calculation of $|\Phi_\Lambda|$ and $|\Phi_g|$ give 14.5 and 1.73 in unit of $10^6\ Km^2 s^{-2}$ respectively, which does not confirm the inequality and the binding state of the SCL2. The large domination of DE (antigravity) over gravity prevent any gravitational influence on the the CrB/A2142sys. Therefore, the analysis can be focalized on the CrB/A2142sys alone, since its inner dynamics largely dominate the whole system.

2) The mass of both the CrB and A2142 superclusters are assumed conentrated in their center of masses. This assumption is giustified by the fact that they have a bound dynamical states at turnaround radii (BP19).

#### 2.1.2. Equations of motion with DE: the Newtonian approximation

The equation of motion in the presence of DE takes the form

$$\frac{d^2r}{dt^2} = -\frac{GM}{r^2} + \frac{8\pi}{3} G\rho_\Lambda r \qquad (1)$$

where r is the distance between the centers of the CrB and A2142 while $M$ is the total mass of the system. The r.h.s. shows the opposite action between the gravitational attraction and the repulsion of DE. The first integral of eq.(1) has the form

$$\frac{1}{2}\left(\frac{dr}{dt}\right)^2 = \frac{GM}{r} + \frac{4\pi}{3} G\rho_\Lambda r^2 + E \qquad (2)$$

where the l.h.s. is the kinetic energy while the first two terms of the r.h.s. define the potential energy and $E$ is the constant of integration corresponding to the total energy of the system which can be easily derived from Eq.(2) since all data are available from BP19 (after correction for the assumed cosmological parametrization), that is:

a) $r = 85.25\ cMpc$ is the separation between the centers of mass of the CrB and A2142.

b) $\frac{dr}{dt} = 4{,}820\ Km\ s^{-1}$ is the relative velocity between the CrB and A2142.

c) $M = 2.74\ x10^{16}{}^{+1.4}_{-1.1}\ M_{sun}$ is the total mass of the CrB/A2142sys obtained by individual mass summation of its members (CrB+A2142+filament) and corrected by bias factor of 1.83 found by Chon et al. (2014)[see BP19]. Note that $M$ is almost comparable with the mass of $2.7\ x10^{16}$ found by S23 for the SCL2. Note that even if they are mass evaluations of the same clustered structure, they are obtained with independent methods which make M a robust mass estimation.

Then $E = -1.47x10^{16} cm^2 s^{-2}$ and the equation for radial dependence on time becomes:

$$\frac{dr}{dt} = \sqrt{\frac{2GM}{r} + \frac{8\pi}{3} G\rho_\Lambda r^2 + 2E} \qquad (3)$$

Now, the boundary between finite and infinite trajectories can be defined by the zeros of the function under the square root of Eq.(3) that is

$F(r) = \frac{2GM}{r} + \frac{8\pi}{3} G\rho_\Lambda r^2 + 2E$ and its derivative $F'(r) = -\frac{2GM}{r^2} + \frac{8\pi G\rho_\Lambda r}{3}$

Since has been assumed that the CrB/A2142sys broke its gravitational binding at a certain time of its past history due to the overcoming of DE on gravity, the decoupling should be happened when the separation relative to their centers of masses became greater than the so-called zero gravity radius $R_\Lambda$ *i.e.* the limiting radius which identifies the spherical surface (centered on one of the two body centers) where gravity and anti-gravity balance each other. After some algebra, $R_\Lambda$ has been derived from $F'(r) = 0$

$$R_\Lambda = \left(\frac{3M}{8\pi\rho_\Lambda}\right)^{1/3} = 33\ cMpc$$

The determination of $R_\Lambda$ allows to establish the minimum energy threshold $E_{min}$ necessary to decouple the CrB/A2142sys. After some algebra, from $F(r) = 0$

$$E_{min} = -\frac{3GM}{2R_\Lambda} = -5.31\ x\ 10^{16}\ cm^2 s^{-2}$$

The condition of decoupling is confirmed since $E > E_{min}$ . Another necessary condition to solve the problem is that the arbitrary initial separation $r_0$ at decoupling must be $r_0 \geq R_\Lambda$. These are the necessery conditions for which the CrB and A2142 can move away from each other in radial motion to infinity.
After the decoupling, $\frac{dr}{dt}$ increases monotonously tending asymptotically to the cosmic expansion velocity given by $V_{exp} = \sqrt{\frac{8\pi}{3} G\rho_\Lambda} r$ as $r \rightarrow \infty$.

Finally, from a second integration of Eq.(3) the time dependence on the distance r is given by

$$t = \int_{r_0}^{r} \frac{dr}{\sqrt{\frac{2GM}{r} + \frac{8\pi}{3} G\rho_\Lambda r^2 + 2E}} \qquad (4)$$

Where *t* is the look-back time at which the hypothesized decoupling happened. Note that Eq.(4) is an elliptic integral of the third kind where solutions are obtained via numerical integration and analytic solutions [see for instance Emelyanov & Kovalyov (2013) and Silbergleit & Chernin (2019)]. Here, the mathematical approach of Bisnovatyi-Kogan & Merafina (2023) has been applied for this specific case where the orbital motion is radial, infinite and null angular momentum (see their § 7).

**3. Results**

The look-back time *t* spent in the relative motion by the CrB/A2142sys from the event of decoupling that began at the separation $r_0 = 33\ cMpc = R_\Lambda$ to the present observed separation $r = 85.25\ cMpc$ has taken $14.7^{+3.6}_{-2.2}$ *Gys* which is a time well beyond the age of the Universe (≈ 13.8 *Gys*). This means that the origin of the CrB/A2142sys, intended as a descendent of a binary system embedded in a galaxy proto-cluster, is not consistent with observations unless the age of the Universe is older than the conventional one as suggested by Gupta (2023). Of course, after such a result the advanced hypothesis on the origins of the CrB/A2142 system can be rejected even if the look-back time at the error lower limit is 12.5 *Gys* ( $z \approx 4.5$) largely consistent with the age and redshift of the most distant galaxy proto-clusters observed in the range of $3 < z < 7$ (Brinch et al. 2023; Li et al. 2025) and beyond at $z \approx$ 9-10 (Wu et al. 2025). It happens because the uncertainty associated to Eq.(4) is very large and essentially ascribable to an error of ~ 50% affecting the estimated total mass *M* of the CrB/A2142sys (BP19) as can be seen in Fig.1.

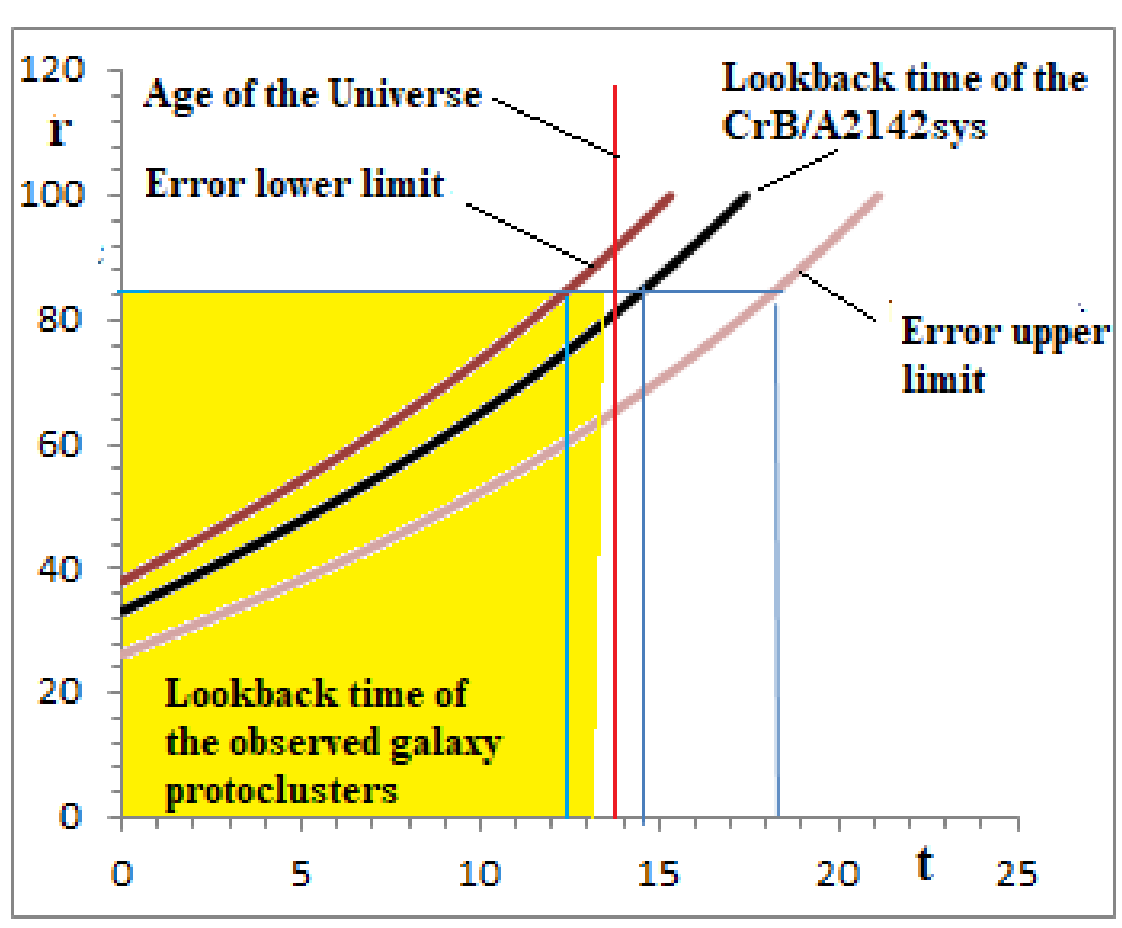

Fig. 1 – Time dependence of the distance for the relative radial motion of the CrB/2142sys. In the x-axis the look-back time $t$ in *Gys* and in the y-axis the separation distance $r$ in *cMpc*. The relative motions are drawn as follows: Main result (black), its error lower limit (purple) and the error upper limit (pink). Note that in the first case the look-back time is greater than the age of the Universe (the red line) while the error lower limit falls within the era of the observed proto-cluster formation ($t < 13.4$ *Gys*; $z < 10$).

## 4. Summary

A recent studies by Nicandro Rosenthal et al. (2025) confirms the observation of a massive galaxy proto-cluster at z~3.14. The peculiarity of this object is the presence of two overdensities at $z = 3.1$ and $z = 3.14$ that are embedded in a large volume of galaxies, a sort of a "Two-in-One system". In the present study has been hypothesized that it could be the primordial-like system from which the CrB/2142sys originated and evolved from. That is, the CrB and A2142, originated from two independent collapsing process within a dense cloud of galaxies. In the context of the ΛCDM model a Newtonian approximation of the two-body motion in presence of DE has been applied using current physical parameters of the CrB/A2142 system with the aim to calculate the look-back time at decoupling. Its compatibility with the era of formation of the primordial galaxy proto-clusters has been tested. The event of the binary decoupling happened at $14.7^{+3.6}_{-2.2}$ *Gys* ago, a time larger than the age of the Universe! However, its compatibility cannot be rejected since the look-back time at the error lower limit is 12.5 *Gys* ( $z \approx 4.5$) largely consistent with observations of the primordial proto-cluster formations.